\documentclass[12pt, prd, showpacs]{revtex4}
\usepackage{amssymb}
\usepackage{amsmath}

\input{tcilatex}

\begin{document}

\title{General limitations on trajectories suitable for super-Penrose process%
}
\author{O. B. Zaslavskii}
\affiliation{Department of Physics and Technology, Kharkov V.N. Karazin National
University, 4 Svoboda Square, Kharkov 61022, Ukraine}
\affiliation{Institute of Mathematics and Mechanics, Kazan Federal University, 18
Kremlyovskaya St., Kazan 420008, Russia}
\email{zaslav@ukr.net }

\begin{abstract}
Collisions of particles near a rotating black hole can lead to unbound
energies $E_{c.m.}$ in their centre of mass frame. There are indications
that the Killing energy of debris at infinity can also be unbound for some
scenarios of collisions near the extremal black hole horizon (so-called
super-Penrose process). They include participation of a particle that (i)
has generic (not fine-tuned) parameters and (ii) moves away from a black
hole before collision. We show that for any finite particle's mass, such a
particle cannot be obtained as a result of the preceding collision. However,
this can be done if one of initial infalling particles has the mass of the
order $N^{-2}$ that generalizes previous observations made in literature for
radial infall in the Kerr background.
\end{abstract}

\maketitle

\section{Introduction}

Nowadays, high energy particles collisions near black holes attract much
attention. In doing so, there are two related but different issues. The
first one consists in the question as to under what conditions high (even
unbound) energies $E_{c.m.}$ in the centre of mass frame can be obtained.
Let the parameters of one of particles be fine-tuned (case (i)). Then, this
can give rise to the so-called Ba\~{n}ados-Silk-West (BSW) effect \cite{ban}%
. In case (ii), no fine-tuning is required \cite{pir1} - \cite{pir3}. There
is also one more case (iii), when fine-tuning of parameters itself is the
consequence of the fact that a particle moved on a circular orbit, provided
this orbit is situated near the horizon of a near-extremal black hole \cite%
{circkerr}, \cite{circ}.

The second issue concerns the properties of debris after collision. The
central question here is whether and under what conditions one can detect at
infinity high Killing energy $E$. The main difficulty here is that a strong
redshift can "eat" even a significant gain of the energy obtained in the
collision. It turned out that the type of scenario between colliding
particles is important here as well. For case (i), when the BSW effect
occurs, it was shown that there exist upper limits on $E$ \cite{p} - \cite{z}%
. Meanwhile, quite recently, a series of works have appeared in which case
(ii) was investigated. It implies that the main effect comes from collision
between an ingoing and outgoing particles near the horizon (head-on
collision, as far as the radial motion is concerned). As was shown in \cite%
{shnit} numerically, this type of scenario can lead to a significant
increase of $E$. In doing so, the outgoing particle was obtained as a result
of reflection of another ingoing particle from the potential barrier. Then,
the outgoing particle under discussion turns out to be close to the
fine-tuned one, provided reflection occurred near the horizon. In \cite{card}%
, an outgoing generic particle was taken as an initial condition \textit{per
se}. As a result, the authors obtained numerically unbound $E$. The latter
property is called in \cite{card} super-Penrose process. It was pointed out
in \cite{pir15} that this leads to some limitations on the mass $m_{\ast }$
of the ingoing particle that should be very large (formally diverging) when
the point of collision approaches the horizon (see also reply in the revised
preprint version of \cite{card}). Head-on collision with participation of a
generic (not fine-tuned) particle of a finite arbirary mass was considered
analytically in \cite{mod}. It was shown there that unbound $E_{c.m.}$ are
indeed possible in such a process.

As the outcome of collision is very sensitive to the type of scenario and
properties of particles before collision, it is important to understand,
what kinds of particles are suitable for the super-Penrose process in
general. The presence of a so-called usual (not fine-tuned) outgoing
particle near the horizon is an essential ingredient for collisions
considered in \cite{card} - \cite{mod}.

In principle, one can impose this initial condition by hand. However, if \ a
particle moves away from the horizon of a black hole, this looks quite
unusual. Therefore, it is desirable to find physical justification for such
trajectories and learn, what such trajectory can come from. More precisely,
we have to elucidate, whether or not they can be obtained from preceding
particle collisions. We consider this issue and show that this is
impossible. The result is valid for any number of particles participating in
the collision near the horizon of the extremal black hole. It is obtained in
a model-independent way, so it does not depend on the particular form of the
metric. Additionally, we generalize the observation made in \cite{pir15} and
obtain the value of $m_{\ast }$ for a quite generic metric and without
additional assumption about radial motion of infalling particles.

It is also worth mentioning that for the extremal Reissner-Nordstr\"{o}m
black hole, unbound $E_{c.m.}$ can be obtained easily even in the standard
BSW picture, without initially outgoing particles \cite{rn}, \cite{esc}.
However, for astrophysical purpose, this metric is irrelevant.

Throughout the paper, the fundamental constants $G=c=1$.

\section{Basic formulas}

We consider the axially symmetric metric of the form%
\begin{equation}
ds^{2}=-N^{2}dt^{2}+g_{\phi }(d\phi -\omega dt)^{2}+\frac{dr^{2}}{A}%
+g_{\theta }d\theta ^{2}\text{,}  \label{met}
\end{equation}%
where the metric coefficients do not depend on $t$ and $\phi $.
Correspondingly, the energy $E=-mu_{0}$ and angular momentum $L=mu_{\phi }$
are conserved. Here, $m$ is a particle's mass, $u^{\mu }=\frac{dx^{\mu }}{%
d\tau }$ being the four-velocity, $\tau $ the proper time along a
trajectory. We restrict ourselves to motion within the equatorial plane $%
\theta =\frac{\pi }{2}$. Then, without the loss of generality, we can
redefine the radial coordinate in such a way that $A=N$. We do not restrict
ourselves to the Kerr or another concrete form of the metric, so the results
are quite general. The equations of motion for a geodesic particle read%
\begin{equation}
m\dot{t}=\frac{X}{N^{2}}\text{,}
\end{equation}%
\begin{equation}
m\dot{\phi}=\frac{L}{g_{\phi }}+\frac{\omega X}{N^{2}}\text{,}
\end{equation}%
\begin{equation}
m\dot{r}=\sigma Z\text{,}  \label{pr}
\end{equation}%
where%
\begin{equation}
X=E-\omega L\text{,}
\end{equation}%
\begin{equation}
Z=\sqrt{X^{2}-N^{2}(m^{2}+\frac{L^{2}}{g_{\phi }})}\text{,}  \label{z}
\end{equation}%
dand the dot denotes differentiation with respect to $\tau $, the factor $%
\sigma =+1$ or $-1$ depending on the direction of motion.

As usual, we assume the forward-in-time, condition $\dot{t}>0$, whence 
\begin{equation}
X\geq 0.  \label{ft}
\end{equation}%
The equality can be achieved on the horizon only where $N=0$. Particles with 
$X_{H}>0$ separated from zero are called usual, particles with $X_{H}=0$ are
called critical. If $X_{H}\neq 0$ but is small we call a particle
near-critical (see next Section for more detailed explanations). Subscript
"H" means that the quantity is calculated on the horizon.

\section{Properties of particle's motion near the horizon.}

In what follows, we consider the extremal horizon. In combination with
formulas from the previous section, this gives rise to some quite definite
properties of motion near the horizon. Let us consider first usual
particles. In the vicinity of the horizon, there is no turning point since $%
N\rightarrow 0$, $X>0$, so $Z^{2}>0$. If the impact parameter $b\equiv \frac{%
L}{E}<\omega _{H}^{-1}$, the particle can reach the horizon. If $b>\omega
_{H}^{-1}$, this is impossible since it would contradict condition (\ref{ft}%
). This means that a usual particle can reach the turning point, where $Z=0$%
, and bounce back. We are interested in collisions with small $N$ only, so
we take into account the situation when the turning point lies close to the
horizon. Then, $b$ must be close to the "critical" value $\omega _{H}^{-1}$.
Near the extremal horizon, the requirement of regularity gives rise to the
expansion \cite{tan}%
\begin{equation}
\omega =\omega _{H}-B_{1}N+O(N^{2})\text{,}  \label{om}
\end{equation}%
where $B_{1}$\ is some model-dependent coefficient. Correspondingly,%
\begin{equation}
X=X_{H}+B_{1}LN+O(N^{2})\text{.}  \label{xn}
\end{equation}

There are two clear definitions of usual ($X_{H}=0$) and critical ($%
X_{H}\neq 0$)\ particles. For near-critical particles, we required $X_{H}$\
to be small (see above) but have not yet specify more precisely, how small $%
X_{H}$\ should be. Now, this can be done on the basis of the expansion (\ref%
{xn}). It is natural to assume that $X_{H}$\ (the value of $X$\ on the
horizon) has the same order as the second term taken in the point of
collision$,$where $N=N_{c}$\ (hereafter, subscript "c" indicates the point
of collision). Otherwise, we return to previous definitions. Indeed, if $%
X_{H}\ll B_{1}LN$, the contribution from $X_{H}$\ is negligible, and the
particle behaves practically like the critical one. The opposite case, $%
X_{H}\gg B_{1}LN$, is typical of a usual particle. Only the intermediate
case when%
\begin{equation}
X_{H}=DN_{c},  \label{nc}
\end{equation}%
deserves special attention. Here, $D$\ is some constant. Thus we specify our
definition of near-critical particles requiring that (\ref{nc}) is valid. In
eq. (\ref{nc}) $N_{c\text{ }}$is not a free variable since it is taken only
in the point of collision. This is in contrast with eqs. (\ref{om}), (\ref%
{xn}) which are valid for arbitrary $N,$\ small enough. Then,%
\begin{equation}
X_{c}=N_{c}(D+B_{1}L)+O(N_{c}^{2})\text{.}
\end{equation}

If the particle is exactly critical, $D=0$. According to our definitions, in
the point of collision both critical and near-critical particles are similar
in the sense that in both cases\textrm{\ }%
\begin{equation}
X_{c}=O(N_{c}).
\end{equation}

\textrm{\ }

It is seen from (\ref{z}) that 
\begin{equation}
Z(N_{c})=O(N_{c}).  \label{zn}
\end{equation}

As a result, for all cases we can write 
\begin{equation}
\lim_{N_{c}\rightarrow 0}Z=X\text{.}  \label{zx}
\end{equation}

In particular, for critical particles both sides of (\ref{zx}) are equal to
zero. If a particle is near-critical, sending $N_{c}\rightarrow 0$\ we also
send $X_{H}$\ to zero since it is adjusted to $N_{c}$\ according to (\ref{nc}%
). In doing so, a near-critical particle becomes more and more similar to
the critical one and, in the limit, turns into it.

\section{Type of outgoing particles from collisions near horizon}

In this section we extend approach of \cite{j} (see also Sec. V of \cite{z})
to the case of multiple collision. Let several particles initially move
toward the horizon and collide in some point near the horizon. We assume
that in the point of collision, the total energy and angular momentum are
conserved. As a consequence, in this point we have%
\begin{equation}
X_{in}=X_{fin}\text{,}  \label{if}
\end{equation}%
where $X_{in}$ is the total contribution of initial particles and $X_{fin}$
is the total contribution of the final outcome. Also, we assume the
conservation of the radial momentum. If there are $p$ particles before
collision and $q$ particles after it, this entails%
\begin{equation}
\dsum\limits_{i=1}^{p}\sigma _{i}Z_{i}=\dsum\limits_{k=1}^{q}\sigma _{k}Z_{k}%
\text{.}  \label{sz}
\end{equation}%
Let us consider the limit in which the point of collision approaches the
horizon. We want to elucidate, whether or not a usual outgoing particle
(that was absent initially) can arise as a result of such a collision. In
doing so, all masses $m_{i}$ and $m_{k}$ are considered to be finite.

\textit{Statement.} If in the initial configuration usual outgoing particles
are absent, they cannot appear after collision.

\textit{Proof. }It follows from (\ref{sz}) that in the limit under
discussion,%
\begin{equation}
\dsum\limits_{i=1}^{n}\sigma _{i}\left( X_{i}\right)
_{H}=\dsum\limits_{k=1}^{m}\sigma _{k}\left( X_{k}\right) _{H}\text{.}
\label{sx}
\end{equation}

By assumption, all initial particles are ingoing. Therefore, in the left
hand side of (\ref{sx}), only terms with $\sigma _{i}=-1$ are present. Then,
we have%
\begin{equation}
-X_{in}=X_{fin}^{(+)}-X_{fin}^{(-)}\text{.}  \label{min}
\end{equation}%
Here, $X_{fin}^{(+)}$ and $X_{fin}^{(-)}$ are, correspondingly,
contributions into $X_{fin}$ coming from outgoing and ingoing particles, 
\begin{equation}
X_{fin}=X_{fin}^{(+)}+X_{fin}^{(-)}\text{.}  \label{tx}
\end{equation}

Now, comparing (\ref{if}), (\ref{min}) and (\ref{tx}), we arrive at the
result%
\begin{equation}
X_{fin}^{(+)}=0\text{.}  \label{f0}
\end{equation}%
However, according to (\ref{ft}), each term in this sum is nonnegative.
Therefore, the equality (\ref{f0}) is possible only in the case in which
each of terms is equal to zero. It means that none of outgoing particles
after collision can be usual which completes the proof. If there are several
critical or near-critical particles initially, the statement still holds
true since such particles do not contribute to $X_{in}$ in the limit under
discussion.

If, instead of taking the exact limit, we consider small but nonzero $N_{c}$%
, we should use eq. (\ref{zn}) instead of eq. (\ref{zx}). Then, instead of (%
\ref{f0}) we obtain that%
\begin{equation}
X_{fin}^{(+)}=O(N_{c})\text{.}
\end{equation}

Thus outgoing particles can be near-critical but, again, they cannot be
usual.

The results can be generalized for nonequatorial motion, provided that $%
\theta $- components of velocities are finite. They will enter $Z,$ being
multiplied by $N^{2}$ \cite{jhep} and will not change the conclusion.

\section{Case of supermassive particle}

If some of particles are so heavy that, formally, $m=O(N^{-1})$ or higher,
eq. (\ref{zx}) becomes invalid and our proof fails. This is just the
situation discussed in \cite{pir15} (see their eq. 4) and \cite{card}. The
authors of Ref. \cite{pir15} considered, under which conditions an outgoing
particle can be obtained near the horizon as a result of collisions of two
particles falling from infinity. It was shown in \cite{pir15} that for the
kerr metric the corresponding mass value of one of particles has the order $%
N^{-2}$. Now, we generalize this observation for an arbitrary metric of the
form (\ref{met}). What is even more important, this enables us to elucidate,
whether or not the main restriction on the mass of the infalling particle
found in \cite{pir15} depends on the details of the process.

Let two particles, $2$ and $\ast $ (we use the same notations as in \cite%
{pir15}), move from infinity (or some finite distance) \ towards the
horizon. After collision, particle $3$ falls in a black hole and particle $X$
escapes to infinity. Then, the conservation of the radial momentum reads%
\begin{equation}
p_{2}^{r}+p_{\ast }^{r}=p_{3}^{r}+p_{X}^{r}\text{.}
\end{equation}%
Here, $p^{r}$\ for each particle is taken from (\ref{pr}), so%
\begin{eqnarray}
p_{2}^{r} &=&-\sqrt{X_{2}^{2}-N^{2}(m_{2}^{2}+\frac{L^{2}}{g_{\phi }})}\text{%
, } \\
p_{\ast }^{r} &=&-\sqrt{(m_{\ast }-\omega _{H}L_{\ast })^{2}-N^{2}(m_{\ast
}^{2}+\frac{L_{\ast }^{2}}{g_{\phi }})}\text{,}
\end{eqnarray}%
\begin{eqnarray}
p_{X}^{r} &=&\sqrt{X_{X}^{2}-N^{2}(m_{X}^{2}+\frac{L_{X}^{2}}{g_{\phi }})}%
\text{, } \\
p_{3}^{r} &=&-\sqrt{(X_{2}+X_{\ast }-X_{X})^{2}-N^{2}(\frac{L_{3}^{2}}{%
g_{\phi }}+m_{3}^{2})}\text{,}
\end{eqnarray}%
where we assume for simplicity that $E_{\ast }=m_{\ast }$ and took into
account the conservation of $X$ (\ref{if}). All particles are taken to be
usual.\ Then, near the horizon, for small $N$, one can write for any
particle that%
\begin{equation}
Z\approx X-\frac{N^{2}}{2X}(m^{2}+\frac{L^{2}}{g_{\phi }})\text{.}
\end{equation}

Then, after some algebra, we obtain in the main approximation that $m_{\ast }
$ should be large,%
\begin{equation}
m_{\ast }=\frac{4X_{X}}{N^{2}}+O(\frac{1}{N})\text{.}  \label{mn}
\end{equation}

For the extremal Kerr metric, $N^{2}\approx \frac{(r-M)^{2}}{4M^{2}}$, where 
$M$ is the black hole mass, $r$ is the Boyer-Lindquiste coordinate. Then,
eq. (\ref{mn}) corresponds just to eq. (4) of \cite{pir15}, if one takes
into account that . In \cite{card}, the large value of $m_{\ast }$ was
related to the simplifying assumption $L_{2}=L_{\ast }=0$ taken in \cite%
{pir15}. However, it follows from our consideration that for any finite
angular momenta the result $m_{\ast }=O(N^{-2})$ is qualitatively the same
and, in the main approximation, depends on $X_{X}$ only.

\section{Discussion and conclusions}

Thus we showed that the most simple option for arranging the super-Penrose
process near black holes is closed. However, there are several other
options. Let us enumerate them. Actually, the present result in combination
with some previous works gives a complete classification of potentially
suitable states.

\subsection{Collision with a supermassive particle}

This includes a very massive particle falling in a black hole in the first
collision to obtain a usual escaping particle in the \ near-horizon region.
This is the inevitable price for it.

\subsection{Collision near past horizons (white holes)}

A particle that moves near the horizon in the outward direction can
correspond to white holes rather than to black ones which suggests one more
type of high energy collisions. \ This circumstance was briefly mentioned in 
\cite{bif}. Detailed consideration of such a process was done in \cite{gpw}.
Now, the particle crosses the horizon with $X_{H}>0$ by assumption, so we
have a usual particle moving in the outward direction. If one bears in mind
the potential astrophysical consequences, there is difficulty here since
white holes are probably unstable (see Sec. 15.2 of \cite{fn}). Nonetheless,
a white hole (past horizon) is an unavoidable part of the whole picture of
an eternal black-white hole. Therefore, the complete and coherent
consideration of all possible scenarios should include this case as well.

\subsection{Collisions with (near)critical particles}

There is a scenario in which one of ingoing near-critical particles bounces
back near the horizon thus turning into an outgoing one and collides
afterwards with another ingoing particle. Then, relatively high energies
were obtained numerically in \cite{shnit}. It remains to be seen
analytically, when this type of scenario indeed leads to the super-Penrose
process.

\begin{acknowledgments}
This work was funded by the subsidy allocated to Kazan Federal University
for the state assignment in the sphere of scientific activities.
\end{acknowledgments}

\end{document}